\documentclass[12pt]{article}
\usepackage{amsmath}
\usepackage{bm}
\usepackage{graphicx}
\begin{document}
\title { Influence of momentum-dependent interactions on balance energy and mass dependence}
\author {Aman D. Sood and  Rajeev K. Puri\\
\it Department of Physics, Panjab University, Chandigarh -160 014,
India.\\} \maketitle
\begin{abstract}
We aim to study the role of momentum-dependent interactions in
transverse flow as well as in its disappearance. For the present
study, central collisions involving mass between 24 and 394 are
considered. We find that momentum-dependent interactions have
different impact in lighter colliding nuclei compared to heavier
colliding nuclei. In lighter nuclei, the contribution of mean
field towards the flow is smaller compared to heavier nuclei where
binary nucleon-nucleon collisions dominate the scene. The
inclusion of momentum-dependent interactions also explains the
energy of vanishing flow in $^{12}C+^{12}C$ reaction which was not
possible with the static equation of state. An excellent agreement
of our theoretical attempt is found for balance energy with
experimental data throughout the periodic table.
\end{abstract}
 Electronic address:~rkpuri@pu.ac.in
\newpage
\section{Introduction}
One of the major goals of heavy-ion collisions at intermediate
energies is to study the properties of hot and dense nuclear
matter formed during a collision. The nuclear equation of state
(EOS) has attracted a lot of attention over the past two decades
because of its wider usefulness in several branches of physics.
However, it is also argued by many authors that the momentum
dependent nature of equation of state can also have a significant
effect in those situations where nuclear matter is mildly and
weakly excited. If matter is highly compressed, the
nucleon-nucleon correlations are already broken due to violent and
frequent nucleon-nucleon collisions. However, if matter is either
weakly or mildly excited, the momentum dependent interactions
(MDI) can have sizeable effects. The initial attempts showed
drastic effects of momentum dependent interactions on collective
flow and particle production \cite{aich87,gale87}. In Ref.
\cite{aich87}, pion yield was found to be suppressed by 30\% once
momentum dependent interactions were included in the evolution of
the reaction. Interestingly, the momentum dependent interactions
also suppressed the nucleon-nucleon collisions by the same amount.
The suppression of nucleon-nucleon collisions due to momentum
dependent interactions happens because of the fact that the
inclusion of momentum dependent interactions accelerates nucleons
in the transverse direction during the initial phase of the
reaction leading to the lower density that results into fewer
nucleon-nucleon collisions. Similarly, if one goes from soft to
hard equation of state, the decrease in pion yield is
$\approx10\%$. Interestingly, a soft equation of state with
momentum dependent interactions (dubbed as SMD) yields the same
transverse momentum as static hard equation of state
\cite{aich87,gale87}. Later on, it was shown in Refs.
\cite{pan93,zhang94} that for asymmetric reactions (eg., Ar+Pb),
the SMD explains the data better than the hard equation of state.
Khoa {\it et al.} \cite{khoa92547} also showed that the inclusion
of momentum dependent interactions leads to the reduced density
and temperature whereas transverse momentum observes reverse
trend. From the above discussion, it is clear that the momentum
dependence of the nuclear mean field plays a major role in
determining the nuclear dynamics and is an important feature for
the fundamental understanding of nuclear matter properties over a
wide range of densities and temperatures. As discussed in detail
in Chap. 3, one has tried to understand the influence of momentum
dependent interactions on flow as well as on its disappearance
using variety of nucleon-nucleon cross sections along with
different static equations of state with an aim to pin down the
nature of equation of state as well as strength of
nucleon-nucleon cross section.\\
However, if we look carefully at the literature, one has often
taken one or two reactions and tried to conclude about the above
cited problems. The need of the hour is to look beyond one or two
reactions. We shall concentrate here on the role of momentum
dependent interactions in collective transverse flow and in its
disappearance in heavy-ion collisions throughout the periodic
table. Moreover, the conclusions of different authors at the same
time are contradictory to each other. For example, Refs. \cite
{zhang94,khoa92547} showed that the flow value using SMD is higher
compared to the hard equation of state for the reactions of
$^{93}$Nb+$^{93}$Nb and $^{40}$Ca+$^{40}$Ca, respectively, at 400
MeV/nucleon, whereas, Ref. \cite {pei89} demonstrated that the
flow value using SMD is smaller compared to hard equation of state
for $^{197}$Au+$^{197}$Au at 400 and 800 MeV/nucleon. References
 \cite{aich87,gale87,pan93,khoa92547} showed (for $^{139}$La+$^{139}$La
at 400 and 800 MeV/nucleon and $^{93}$Nb+$^{93}$Nb at 400
MeV/nucleon), that flow values are same for the SMD and hard
equation of state. On the contrary, Refs. \cite {mota92,zhou94n}
showed that the flow is insensitive to the equation of state
irrespective of the momentum dependent interactions. Similarly,
Ref. \cite {khoa92547} demonstrated (for $^{40}$Ca+$^{40}$Ca at
400 MeV/nucleon) that the soft equation of state gives larger flow
compared to hard equation of state whereas Refs.
\cite{khoa92547,pei89,blat91} showed (for $^{197}$Au+$^{197}$Au at
200-800 MeV/nucleon, $^{93}$Nb+$^{93}$Nb at 400 MeV/nucleon, and
$^{139}$La+$^{139}$La at 800 MeV/nucleon, respectively) that a
soft equation of state gives smaller flow. Reference \cite
{zhang94} demonstrated better agreement with the data using SMD
for Ar+Pb reaction whereas Ref. \cite {pei89} concluded just the
opposite for $^{197}$Au+$^{197}$Au reaction. The balance energy
($E_{bal}$), (the incident energy at which the transverse flow
disappears) using static and momentum dependent interactions, also
faces similar contradictions \cite {sull90,krof91,he96}. All the
above mentioned examples indicate a need for the systematic study
of the influence of momentum dependent interactions on flow as
well as on its disappearance throughout the periodic table to pin
down the above mentioned questions and universal behaviour of
momentum dependent interactions.\\

It is worth mentioning that the mass dependence studies have been
performed in a variety of problems in heavy-ion physics. For
example, the study of mass dependence in the evolution of density
and temperature reveals that maximum density scales with the size
of the system \cite {khoa92547,blat91,hart} whereas maximum
temperature is insensitive towards the mass of the system.
Similarly, multifragmentation, particle production, and collective
flow (the most sensitive observable) also depend strongly on the
mass of the system \cite {hart}.\\

It has been discussed by several authors that the above mentioned
observables are mass dependent and the agreement/disagreement with
experimental observations depends upon the size of the reacting
partner. In Chap. 3, we presented a complete theoretical analysis
of balance energy using quantum molecular dynamics model where
model was confronted against balance energy observed in
$^{20}$Ne+$^{27}$Al \cite {west93}, $^{36}$Ar+$^{27}$Al \cite
{ang97,buta95}, $^{40}$Ar+$^{27}$Al \cite {sull90},
$^{40}$Ar+$^{45}$Sc \cite {west93,pak96,mag0062},
$^{40}$Ar+$^{51}V$ \cite {krof91,ogi90}, $^{64}$Zn+$^{27}$Al \cite
{he96}, $^{40}$Ar+$^{58}$Ni \cite {cus02}, $^{64}$Zn+$^{48}$Ti
\cite {buta95}, $^{58}$Ni+$^{58}$Ni \cite
{mag0062,cus02,west01,pak97}, $^{58}$Fe+$^{58}$Fe
\cite{west01,pak97}, $^{64}$Zn+$^{58}$Ni \cite {buta95},
$^{86}$Kr+$^{93}$Nb \cite {west93,mag0062}, $^{93}$Nb+$^{93}$Nb
\cite {krof92}, $^{129}$Xe+$^{118}$Sn \cite {cus02},
$^{139}$La+$^{139}$La \cite {krof92}, and $^{197}$Au+$^{197}$Au
systems \cite {mag0062,zhang90,mag0061}. Apart from the above
mentioned and extensively analyzed reactions, $^{12}$C+$^{12}$C,
though, has also been subjected to balance energy, interestingly,
is the content out of the discussion \cite {west93}. Therefore,
one also needs to understand the dynamics involved in the balance
energy of $^{12}$C+$^{12}$C system. As mentioned above, some
attempts are made in literature where equation of state with
momentum dependent interactions is employed in heavy-ion
collisions to study the balance energy \cite {zhou94,leh96}.
However, very few attempts yet exist where mass dependence of the
balance energy is reported in the literature using momentum
dependent interactions \cite{zhou94}. The mass dependence using
momentum dependent interactions may have interesting physics
since, the surface contribution in lighter nuclei is much larger
compared to the heavier nuclei. For example, the ratio of the
surface to radius is 0.12 in $^{12}$C+$^{12}$C system whereas it
is 0.022 in $^{197}$Au+$^{197}$Au system (the surface is defined
as the radial distance marked with density between 90\% and 10\%
of its central value whereas radius is the distance where density
falls to 50\% of the central density). In other words, the surface
effects and surface to volume ratio will be much stronger in light
nuclei
compared to heavy ones.\\

Our aim, in the present chapter, therefore, is at least twofold.

(1) To study the $^{12}$C+$^{12}$C system for the collective
transverse flow and its disappearance.

(1) To understand the role of momentum dependent interactions in
the collective transverse flow as well as in its disappearance
throughout the periodic table in central collisions and to analyze
whether mass dependence can be presented in terms of some scaling
relation or not.\\
\section{The model}

We simulate the nucleons within the framework of quantum molecular
dynamics (QMD) model. In the QMD model,\cite{aich91,stoc86} each
nucleon propagates under the influence of mutual interactions. The
propagation is governed by the classical equations of motion:
\begin{equation}
\dot{{\bf r}}_i~=~\frac{\partial H}{\partial{\bf p}_i}; ~\dot{{\bf
p}}_i~=~-\frac{\partial H}{\partial{\bf r}_i},
\end{equation}
where \emph{H} stands for the Hamiltonian which is given by:
\begin{equation}
H = \sum_i^{A} {\frac{{\bf p}_i^2}{2m_i}} + \sum_i^{A}
({V_i^{Skyrme} + V_i^{Yuk} + V_i^{Coul}+V_i^{mdi}}).
\end{equation}
The $V_{i}^{Skyrme}$, $V_{i}^{Yuk}$, $V_{i}^{Coul}$, and
$V_i^{Cmdi}$ in Eq. (2) are, respectively, the Skyrme, Yukawa,
Coulomb and momentum-dependent potentials.

The momentum dependent interactions are obtained by parameterizing
the term taken from the measured energy dependence of the
nucleon-nucleus optical potential. It can be parameterized as
\begin{equation}
V_{ij}^{MDI}=t_{4}ln^{2}[t_{5}({\bf p_{i}}-{\bf
p_{j}})^{2}+1]\delta({\bf r_{i}}-{\bf r_{j}}).
\end{equation}
Here $t_{4}$ = 1.57 MeV and $t_{5}$ = $5\times 10^{-4}$
MeV$^{-2}$. The final form of the momentum dependent potential
reads as
\begin{equation}
U^{MDI}=\delta .ln^{2}[\epsilon(\rho / \rho_{0})^{2/3}+1]\rho /
\rho_{0}.
\end{equation}
 A parameterized form of the local plus momentum dependent
potential is given by
\begin{equation}
U=\alpha \left({\frac {\rho}{\rho_{0}}}\right) + \beta
\left({\frac {\rho}{\rho_{0}}}\right)^{\gamma}+ \delta
ln^{2}[\epsilon(\rho/\rho_{0})^{2/3}+1]\rho/\rho_{0}.
\end{equation}
It can be used to compute the corresponding density dependence of
the compressional energy per nucleon which is shown in Fig.
\ref{eos} for the soft (dubbed as Soft) and hard (dubbed as Hard)
local Skyrme potentials and for the interactions with a momentum
dependent term which is denoted by SMD and HMD.

\section{Results and Discussion}
\hspace*{0.5cm} We simulated 1000-3000 events for each of
$^{12}$C+$^{12}$C ($b/b_{max}=0.4$), $^{20}$Ne+$^{27}$Al
($b/b_{max}=0.4$), $^{36}$Ar+$^{27}$Al ($b=2$ fm),
$^{40}$Ar+$^{27}$Al ($b=1.6$ fm), $^{40}$Ar+$^{45}$Sc
($b/b_{max}=0.4$), $^{40}$Ar+$^{51}$V ($b/b_{max}=0.3$),
$^{40}$Ar+$^{58}$Ni ($b$ = 0-3 fm), $^{64}$Zn+$^{48}$Ti ($b=2$
fm), $^{58}$Ni+$^{58}$Ni ($b/b_{max}=0.28$), $^{64}$Zn+$^{58}$Ni
($b=2$ fm), $^{86}$Kr+$^{93}$Nb ($b/b_{max}=0.4$),
$^{93}$Nb+$^{93}$Nb ($b/b_{max}=0.3$), $^{129}$Xe+$^{Nat}$Sn ($b$
= 0-3 fm), $^{139}$La+$^{139}$La ($b/b_{max}=0.3$), and
$^{197}$Au+$^{197}$Au ($b=2.5$ fm) using HMD and hard equations of
state at various incident energies between 30 MeV/nucleon and 800
MeV/nucleon in small steps. The constant cross sections of 40, 50,
and 55 mb strength are used at each such incident energy.
 The
impact parameters are taken from the experimental extractions
\cite{sull90,krof91,he96,west93,ang97,buta95,pak96,mag0062,ogi90,cus02,west01,pak97,krof92,zhang90,mag0061}.
The directed transverse momentum is calculated using $\langle
p_{x}^{dir}\rangle$ \cite {leh96,aich91,sood04,kum98,hart98}
\begin{equation}
\langle p_{x}^{dir}\rangle=\frac{1}{A}\sum_i {\rm
sgn}\{Y(i)\}p_{x}(i),
\end{equation}
where $Y(i)$ and $p_{x}(i)$ are, respectively, the rapidity
distribution and transverse momentum of {\it i}th particle.\\
\begin{figure}[!t]
\centering
 \vskip -3cm
\includegraphics[width=13cm]{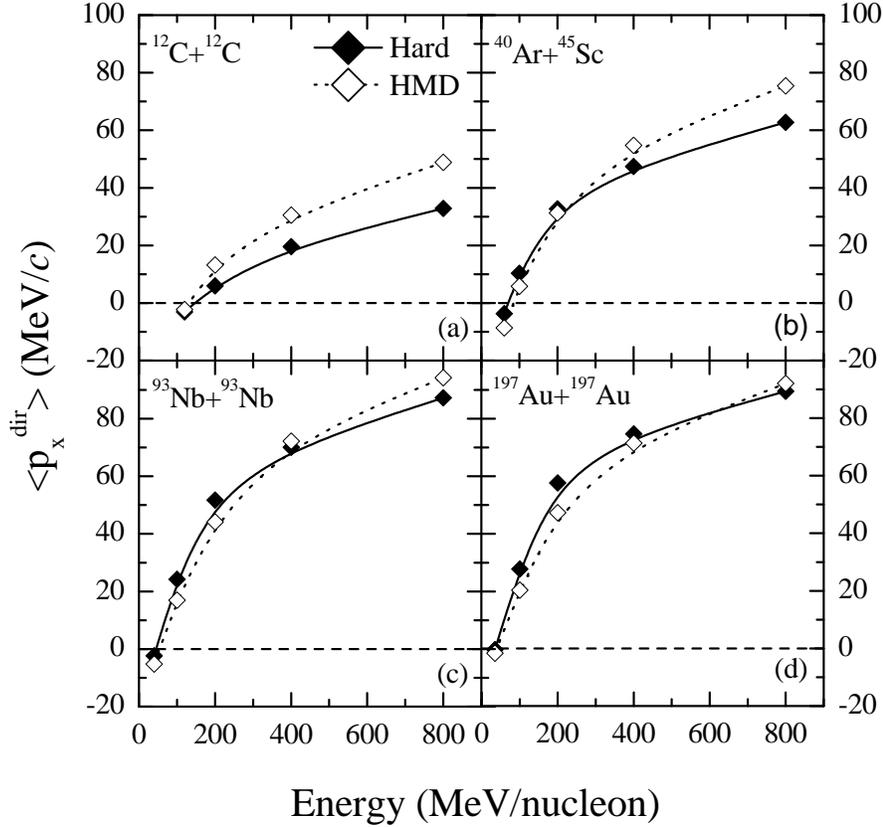}
 \vskip -5cm
\caption{The $\langle p_{x}^{dir} \rangle$ as a function of
incident energy for four different reactions. (a)
$^{12}$C+$^{12}$C, (b) $^{40}$Ar+$^{45}$Sc, (c)
$^{93}$Nb+$^{93}$Nb, and (d) $^{197}$Au+$^{197}$Au. Solid (open)
diamonds represent hard (HMD) equation of state.  The lines are
only to guide the eye. A constant nucleon-nucleon cross section of
55 mb strength is used.}\label{4.1}
\end{figure}
\begin{figure}[!t]
\centering \vskip -3cm
\includegraphics[width=13cm]{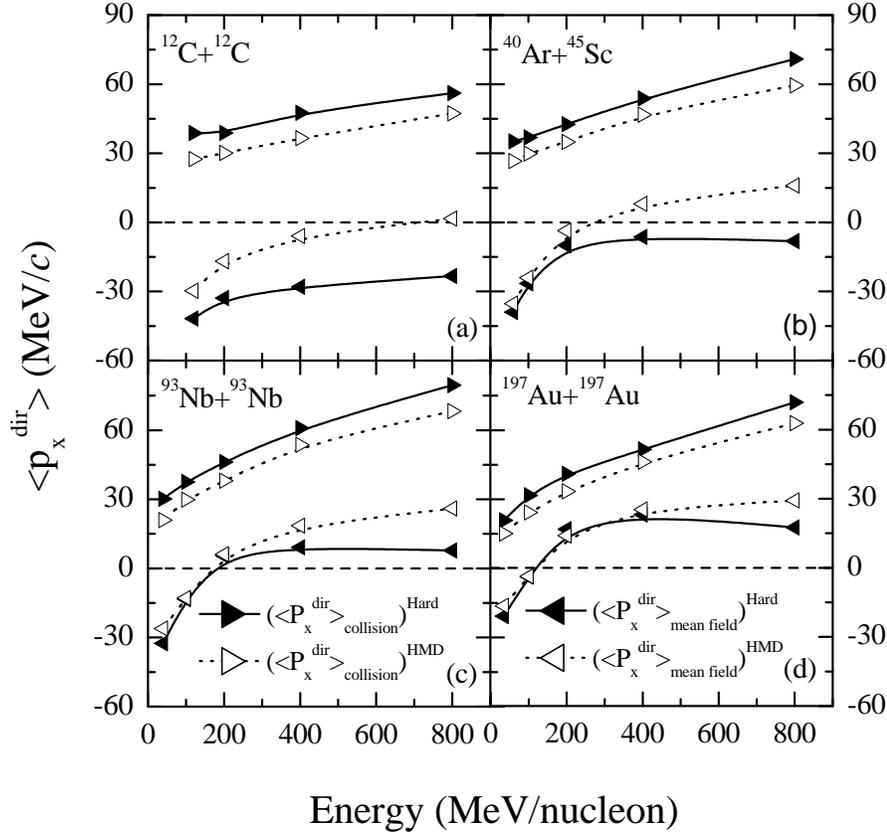}
 \vskip -5cm
\caption{The decomposition of $\langle p_{x}^{dir} \rangle$ into
mean field (left triangles) and collision part (right triangles)
as a function of incident energy. Again, solid (open) triangles
represent hard (HMD) equation of state.}\label{4.2}
\end{figure}
\begin{figure}[!t]
\centering
 \vskip -3cm
\includegraphics[width=10cm]{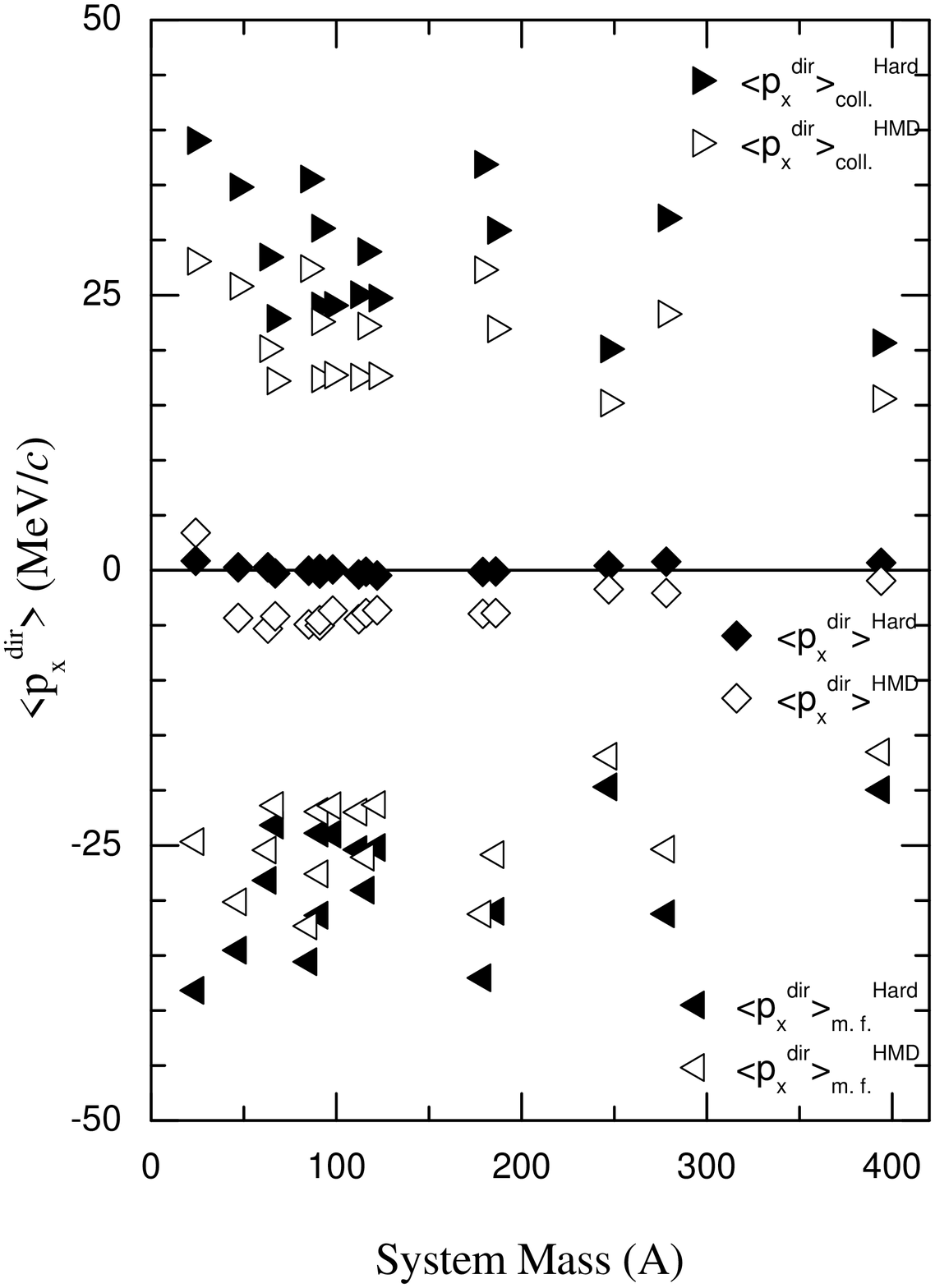}
 \vskip -0.2cm
\caption{Total $\langle p_{x}^{dir} \rangle$ (diamonds) and its
decomposition into mean field (left triangles) and collision part
(right triangles) as a function of the mass of the system at the
balance energy of hard equation of state. Solid (open) symbols
represent hard (HMD) equation of state.}\label{4.3}
\end{figure}\\
\hspace*{0.5cm}In Fig. \ref{4.1}, we display $\langle p_{x}^{dir}
\rangle$ as a function of incident energy ranging between 30
MeV/nucleon and 800 MeV/nucleon, for the reactions of
$^{12}$C+$^{12}$C, $^{40}$Ar+$^{45}$Sc, $^{93}$Nb+$^{93}$Nb, and
$^{197}$Au+$^{197}$Au. The open (solid) diamonds denote the
$\langle p_{x}^{dir}\rangle$ values for HMD (Hard) equation of
state. A constant cross section of 55 mb has been used in this
figure. The lines are to guide the eye. In all the cases,
transverse momentum is negative at lower incident energies which
turns positive at relatively higher incident energies. The value
of the abscissa at zero value of $\langle p_{x}^{dir}\rangle$
corresponds to the energy of vanishing flow (EVF) or,
alternatively, the balance energy ($E_{bal}$). The following
important results emerge from the graph.
\begin{figure}[!t]
\centering
 \vskip -2cm
\includegraphics[width=10cm]{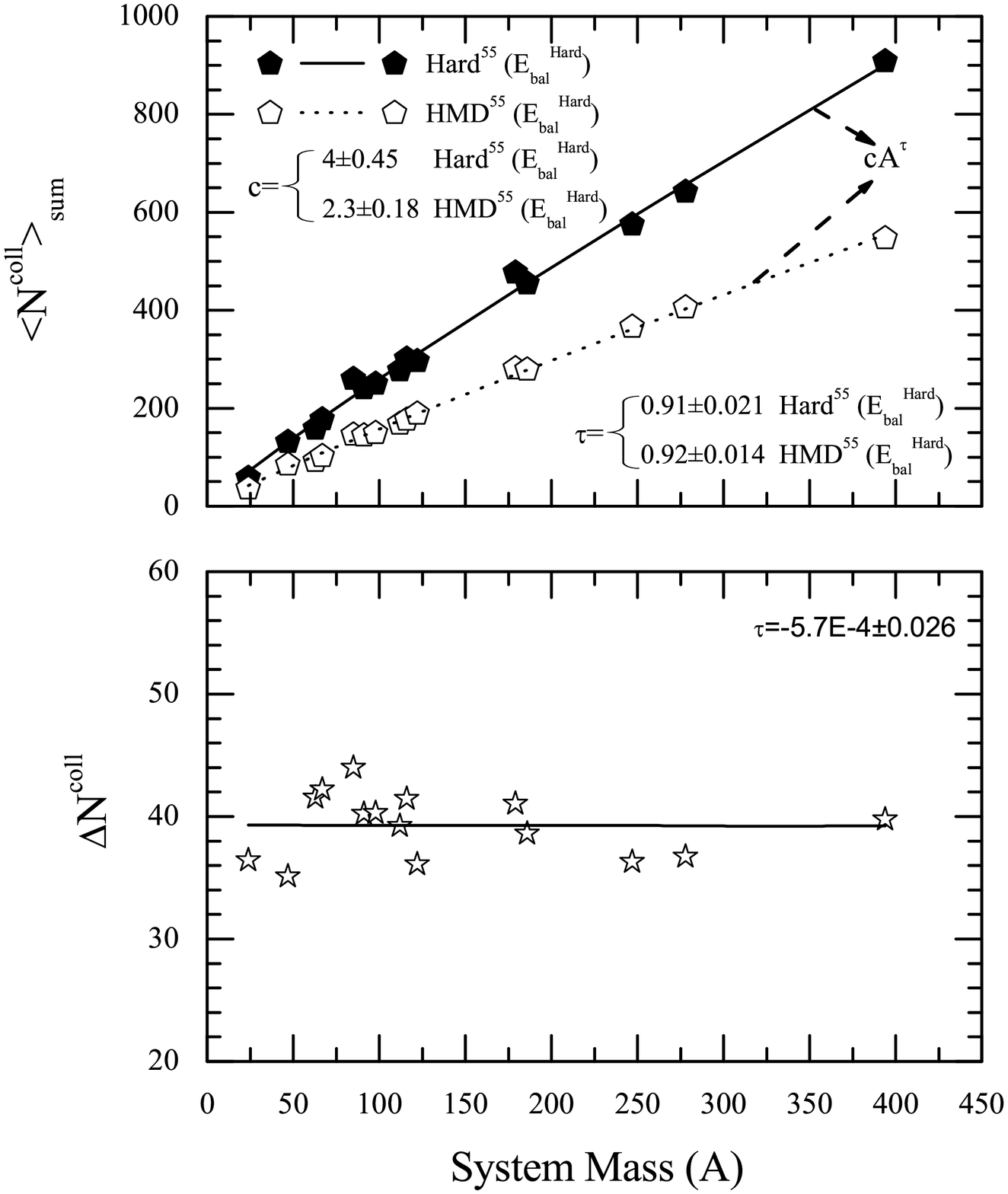}
 \vskip 0.4cm
\caption{(a) The total number of allowed nucleon-nucleon
collisions as a function of total mass of the system at the
balance energy of hard equation of state. Solid (open) pentagons
represent hard (HMD) equation of state. The lines are a power law
fit of form $cA^{\tau}$. Solid (dotted) line represents power law
fit for hard (HMD) equation of state. (b) The percentage
difference $\Delta N^{coll}$ as a function of mass of the
system}\label{4.5}
\end{figure}
\begin{figure}[!t]
\centering
 \vskip -3cm
\includegraphics[width=13cm]{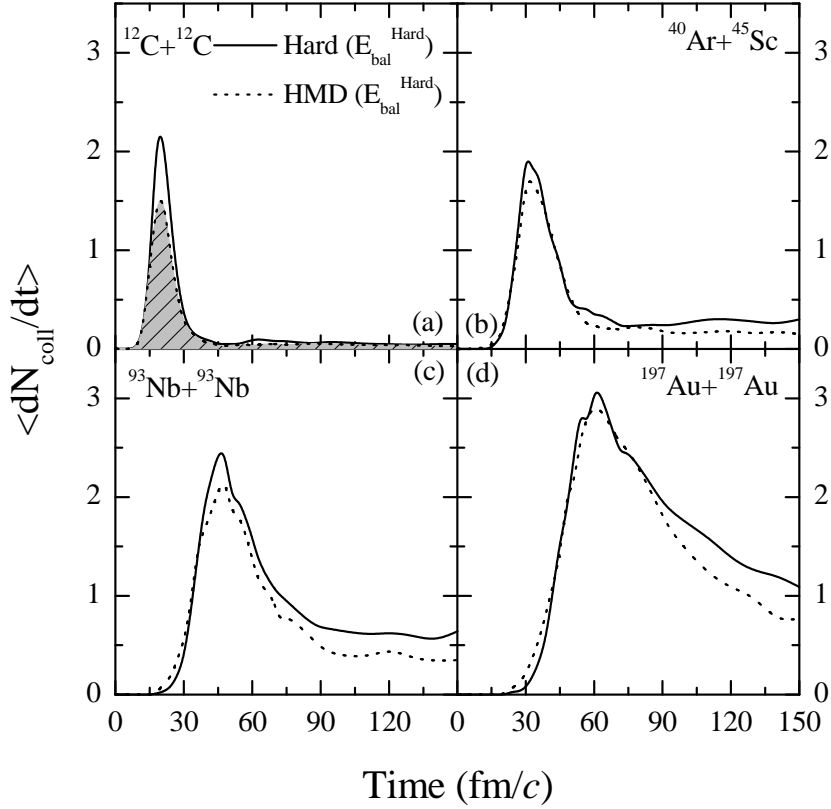}
 \vskip -5cm
\caption{The rate of allowed nucleon-nucleon collisions
$dN_{coll}/dt$ versus reaction time for the reactions of (a)
$^{12}$C+$^{12}$C, (b) $^{40}$Ar+$^{45}$Sc, (c)
$^{93}$Nb+$^{93}$Nb, and (d) $^{197}$Au+$^{197}$Au. Solid (dotted)
lines represent hard (HMD) equation of state.}\label{4.6}
\end{figure}
\begin{figure}[!t]
\centering
 \vskip -3cm
\includegraphics[width=13cm]{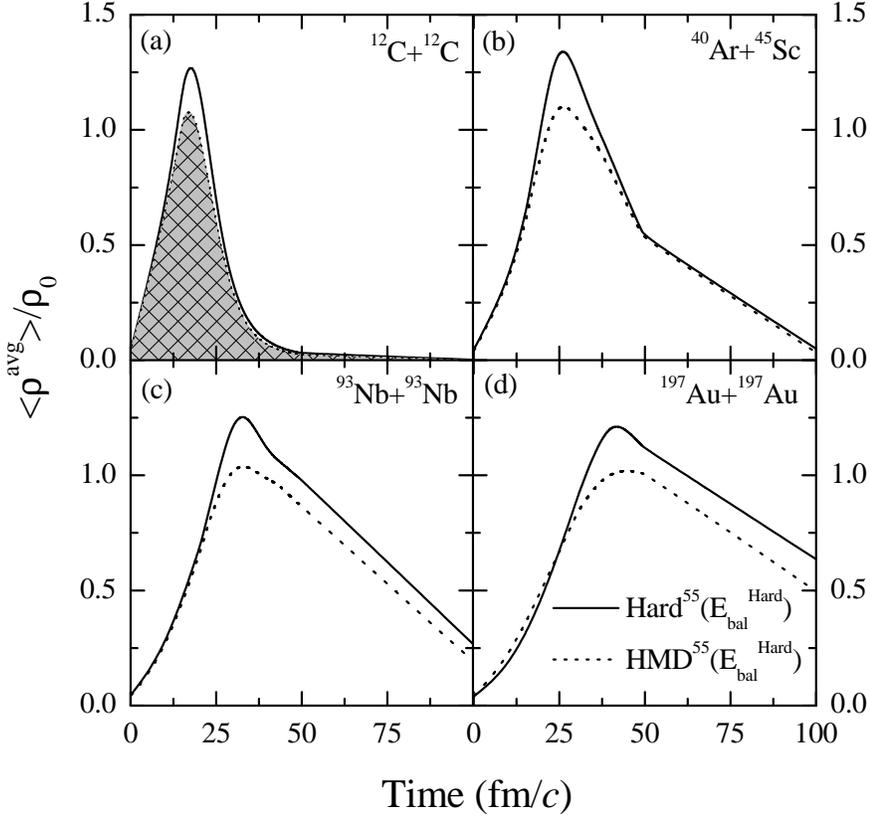}
 \vskip -5cm
\caption{ Same as Fig. \ref{4.6}, but average density ($\langle
\rho^{avg} \rangle /\rho_{0}$) versus reaction time.}\label{4.7}
\end{figure}
\begin{figure}[t]
\centering
 \vskip -3cm
\includegraphics[width=8cm]{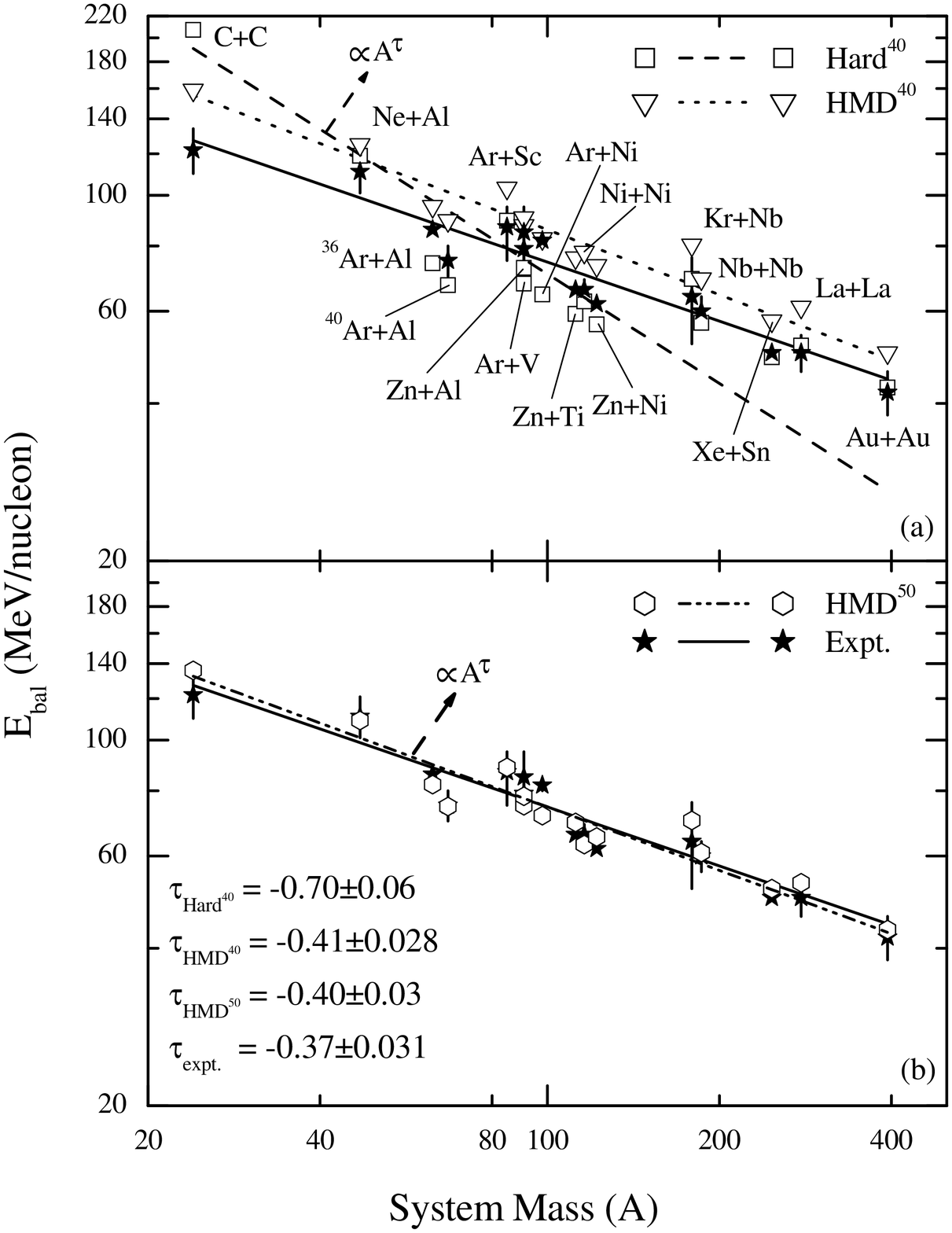}
 \vskip -0.2cm
\caption{(a) The balance energy as a function of combined mass of
the system. The experimental points along with error bars are
displayed by solid stars. Our calculations for $\sigma=40$ mb with
hard (HMD) equation of state are represented by open squares (open
down triangles). (b) Same as Fig. \ref{4.8}a, but using HMD with
$\sigma=50$ mb strength (denoted by open hexagons) and
experimental data (denoted by solid stars). The lines are power
law fit of the form $cA^{\tau}$. The dashed, dotted, and
dash-double-dotted lines are power law fit for Hard$^{40}$,
HMD$^{40}$, and HMD$^{50}$, respectively, whereas solid line
represents experimental points.}\label{4.8}
\end{figure}
(a) The transverse momentum increases monotonically with increase
in the incident energy. The increase in transverse flow $\langle
p_{x}^{dir}\rangle$ is sharp at smaller incident energies (up to
200 MeV/nucleon) compared to higher incident energies where it
starts saturating. Its slope decreases at higher incident energies
that finally saturates depending upon the mass of the colliding
nuclei.

(2) At higher energies (eg., above 400 MeV/nucleon), the repulsion
due to momentum dependent interactions is stronger during the
early phase of reaction and transverse momentum increases sharply.
However, the overall effect depends on the mass of colliding
nuclei. The difference of $\langle p_{x}^{dir}\rangle$ between HMD
and hard equations of state decreases as one goes from lighter to
heavier systems. For example, the difference ($\langle
p_{x}^{dir}\rangle_{HMD}$-$\langle p_{x}^{dir}\rangle_{Hard}$) at
400 MeV/nucleon is approximately, 11, 7.3, 2, and -3.2 MeV/{\it
c}, respectively, for the reactions of $^{12}$C+$^{12}$C,
$^{40}$Ar+$^{45}$Sc, $^{93}$Nb+$^{93}$Nb, and
$^{197}$Au+$^{197}$Au. If one calculates in terms of normalized
percentage [i.e., \{($\langle p_{x}^{dir}\rangle_{HMD}$-$\langle
p_{x}^{dir}\rangle_{Hard}$)/$\langle
p_{x}^{dir}\rangle_{Hard}$\}$\times$100], these numbers are
modified to 56.35, 15.44, 3.04, and -4.31\%, respectively, for the
reactions of $^{12}$C+$^{12}$C, $^{40}$Ar+$^{45}$Sc,
$^{93}$Nb+$^{93}$Nb, and $^{197}$Au+$^{197}$Au. Note that the
lighter colliding nuclei now show a huge variation compared to
heavy ones where effects are insignificant. (3) If one looks
carefully at the reactions reported in Fig. \ref{4.1} in the
vicinity of balance energy, one sees that the momentum dependent
interactions suppress the transverse momentum and hence enhance
the balance energy in agreement with the results of \cite
{soff95,leh96}. The enhancement in the balance energy due to
momentum dependent interactions over static one is, respectively,
13.29, 10.11, and 2.6 MeV/nucleon for the reactions of
$^{40}$Ar+$^{45}$Sc, $^{93}$Nb+$^{93}$Nb, and
$^{197}$Au+$^{197}$Au. On the contrary, momentum dependent
interactions reduce the energy of vanishing flow in
$^{12}$C+$^{12}$C by 9.9 MeV/nucleon.
All the above mentioned findings can be understood by decomposing
the total transverse momentum into contributions due to mean field
and two-body nucleon-nucleon collisions (as shown in Fig.
\ref{4.2}). This decomposition has been explained in details in
Ref. \cite{sood04}. The left triangles represent mean field
contribution whereas collision contribution is displayed by the
right triangles. The open (solid) triangles represent HMD (Hard)
equation of state. One notices that the flow due to binary
nucleon-nucleon collisions increases almost linearly with increase
in the incident energy in all the above listed reactions. The mean
field flow, however, increases sharply up to couple of hundred
MeV/nucleon then saturates. This trend is slower in lighter
colliding nuclei compared to heavy ones. In other words, the
continuous increase in the total transverse momentum in high
energy tail results due to the frequent nucleon-nucleon binary
collisions alone. Under this saturation, the overall trend of the
role of momentum dependent interactions in transverse flow is
decided by the impact of the mean field. The striking result is
for the case of $^{12}$C+$^{12}$C reaction, where the contribution
of the mean field towards collective transverse flow using
momentum dependent interactions dominates the decrease in the
collective flow due to nucleon-nucleon collisions using momentum
dependent interactions resulting in the net increased transverse
flow due to HMD compared to the static hard equation of state.
Whereas, for medium mass nuclei (e.g., $^{93}$Nb+$^{93}$Nb), the
increase in the flow due to mean field is almost balanced by the
decrease in flow due to nucleon-nucleon binary collisions,
therefore neutralizing the net effect. For the heavier systems, a
very little swing can be noticed. In other words, the role of the
momentum dependent interactions in mean field contribution,
depends on the mass of the system. In lighter systems, the role of
momentum dependent interactions is larger which decreases for the
heavier colliding nuclei.
\hspace*{0.5cm} Let us now examine the mass dependence of momentum
dependent interactions in balance energy$\rightarrow$ A point in
the energy scale corresponding to vanishing flow. A careful look
at Fig. \ref{4.2} shows that the balance energy is, respectively,
142.88 (133), 70.11 (83.4), 47.18 (57.29), and 37.44 (40.1)
MeV/nucleon for the reactions of $^{12}$C+$^{12}$C,
$^{40}$Ar+$^{45}$Sc, $^{93}$Nb+$^{93}$Nb, and
$^{197}$Au+$^{197}$Au using hard (HMD) equation of state. All the
reactions, except $^{12}$C+$^{12}$C, show a uniform trend, i.e.,
the inclusion of momentum dependent interactions reduces the
transverse flow at low incident energies, therefore pushing the
balance energy towards higher end. However, $^{12}$C+$^{12}$C
shows just the opposite. To understand this, let us divide the
total transverse momentum at the energy of vanishing flow into
contributions resulting from the mean field as well as collision
parts. In Fig. \ref{4.3}, we plot the decomposition for all the
reactions reported in the introduction where balance energy has
been measured and reported experimentally. As discussed above, the
difference between the transverse flow contributions due to mean
field of HMD and hard equations of state is maximal for lighter
colliding nuclei which gets suppressed for the heavier colliding
nuclei.
It was argued by Zhou {\it et al.} \cite {zhou94} that for the
lighter systems like $^{12}$C+$^{12}$C, a larger value of the
balance energy results in stronger momentum dependent repulsion
that enhances the transverse momentum and hence suppresses the
balance energy. Our findings are also pointing towards the same
effects. However, one should also keep in mind that this trend is
not universal in the mass range. Looking at Fig. \ref{4.1}, one
notices that the collisions of heavier colliding nuclei at the
balance energy of $^{12}$C+$^{12}$C reaction do not yield any
significant difference or trend shown by the $^{12}$C+$^{12}$C
reaction.
Let us check whether the inclusion of momentum dependent
interactions reduces the frequency of nucleon-nucleon binary
collisions or not. We display in Fig. \ref{4.5}(a), the total
number of nucleon-nucleon collisions observed in the simulations
using HMD and hard equations of state as a function of the mass of
the system at their corresponding theoretical balance energy for a
hard equation of state. The aim to simulate the reactions of HMD
also at the balance energy of static hard equation of state is to
eliminate any variation that might result due to different balance
energy in HMD compared to hard equation of state. The solid line
corresponds to hard equation of state whereas dotted line is
showing the outcome for the HMD equation of state. As is evident,
the inclusion of momentum dependent interactions suppresses the
binary collisions by as much as 35-45\%  throughout the mass range
which is in close agreement with Ref. \cite {aich87}. Further,
these can be parameterized by a power law of the form $cA^{\tau}$
with $\tau=0.92\pm0.014$ and $\tau=0.91{\pm}0.021$, respectively,
for the HMD and hard equations of state. The similar values of
power law parameter $\tau$ indicate towards universal suppression
of binary collisions due to inclusion of momentum dependent
interaction. To look more carefully, we plot in Fig. \ref{4.5}(b),
the percentage difference of binary nucleon-nucleon collisions
defined as
\begin{equation}
\Delta{N^{coll}}=|\frac{<N^{coll}>_{HMD}-<N^{coll}>_{Hard}}{<N^{coll}>_{Hard}}|\times
{100}. \nonumber
\end{equation}
As stated above, we see that the average suppression is of the
order of 40\% throughout the periodic table masses. Since above
discussion was for the final stage collisions, it will be of
further interest to see how collisions are affected during the
course of the reaction. To see how momentum dependent interactions
can affect the frequency of the binary collisions, we show in Fig.
\ref{4.6}, the rate of change of allowed collisions dN/dt as a
function of time for $^{12}$C+$^{12}$C, $^{40}$Ar+$^{45}$Sc,
$^{93}$Nb+$^{93}$Nb, and $^{197}$Au+$^{197}$Au. This rate is after
eliminating the Pauli blocked collisions. Interestingly, during
early phase of the reaction, the inclusion of momentum dependent
interactions has drastic effect on the collision rate compared to
hard equation of state in lighter colliding nuclei. This trend is
reversed in the case of heavier nuclei where very little effect
can be seen. This also points towards the stronger effect of
momentum dependent interactions in lighter nuclei compared to
heavy colliding nuclei. To understand the suppression of the flow
using momentum dependent interactions, we display in Fig.
\ref{4.7}, the time evolution of the rescaled density, for the
reactions of $^{12}$C+$^{12}$C, $^{40}$Ar+$^{45}$Sc,
$^{93}$Nb+$^{93}$Nb, and $^{197}$Au+$^{197}$Au at the balance
energy for hard equation of state (solid line) and HMD (dotted
line). One notices that the momentum dependent interactions
suppress the high dense phase of the reaction even at low incident
energies like the balance energy. The reduction in the maximal
value of the average density indicates a reduction in the number
of nucleon-nucleon collisions. Note that the reduction in the
maximal value of the average density is nearly the same in all the
reactions irrespective of quite different energy of vanishing flow
indicating nearly the same reduction in number of binary
nucleon-nucleon collisions and hence collision transverse flow for
all the reacting systems. One also notices a faster decomposition
of the compressed system using momentum dependent interactions as
has been reported in Ref. \cite {blat91}.
 In our previous works \cite
{sood04}, we had reported the mass dependence of the disappearance
of transverse flow for the whole range ranging from
($^{20}$Ne+$^{27}$Al) to heavier masses ($^{238}U$+$^{238}U$). Our
comparison with experimental data suggested a preference for the
hard equation of state. Also nucleon-nucleon cross section of
35-40 mb explained the data throughout the mass range quite
nicely. The experimental balance energy yielded
$\tau_{expt}=-0.42\pm0.05$ whereas our results predicted
$\tau_{35}=-0.43\pm0.09$ and $\tau_{40}=-0.42\pm0.08$ for the
cross sections of 35 and 40 mb, respectively. We here extend the
above study to also include the $^{12}$C+$^{12}$C reaction,
therefore, increasing the mass range from 24 to 394. In addition,
we shall also study the role of momentum dependent interactions.
In Fig. \ref{4.8} we display the balance energy as a function of
the total mass of the system from $^{12}$C+$^{12}$C to
$^{197}$Au+$^{197}$Au. We display the results for a hard equation
of state (open square) and HMD (inverted triangles) with
$\sigma=40$ mb. An enhanced nucleon-nucleon cross section with
$\sigma=50$ mb is also used in the case of momentum dependent
interactions (open hexagon). Experimental data are displayed by
stars. The lines are the power law fits of the form ($\propto
A^{\tau}$). The dashed, dotted, dash-double-dotted, and solid
lines represent, respectively, the power law fit for Hard$^{40}$,
HMD$^{40}$, HMD$^{50}$, and experimental data. The value of $\tau$
for experimental data is $\tau_{expt}=-0.37\pm0.031$ whereas
$\tau_{Hard^{40}}=-0.7\pm0.06$, $\tau_{HMD^{40}}=-0.38\pm0.029$,
and $\tau_{HMD^{50}}=-0.4\pm0.003$. Once momentum dependent
interactions are included, value of the $\tau_{HMD^{40}}$ comes
very close to the $\tau_{expt.}$. As noted in upper part of the
figure, inclusion of momentum dependent interactions improves the
agreement for $^{12}$C+$^{12}$C system. One still notices that in
most of the medium and heavy masses, it rather overestimates the
balance energy. To improve the agreement further, we also
simulated the reactions at $\sigma=50$ mb using momentum dependent
interactions. The comparison is displayed in the lower part of the
figure. We see excellent agreement of HMD$^{50}$ with experimental
balance energy throughout the periodic table with mass between 24
and 394. Obviously, the effect is larger for lighter nuclei where
incident energy is higher compared to heavy nuclei having lower
incident energies. The failure of hard equation of state is due to
the fact that it fails badly to explain the data for
$^{12}$C+$^{12}$C reaction. It seems that the static equation of
state is not able to generate enough repulsion in terms of
transverse momentum in lighter colliding nuclei. Though as
reported in Ref. \cite{sood04}, other balance energies can be
nicely explained. One possibility is to also use enhanced cross
section for the static hard equation of state. This, however
worsen the balance energy agreement for medium and heavy nuclei.

\section{Summary}

We have studied the role of momentum dependent interactions in
transverse flow as well as in its disappearance for central
collisions over a wide range of masses between 24 and 394. We find
that even for the central collisions at low incident energies, the
role of momentum dependent interactions is significant and is not
uniform over the mass range. The impact of momentum dependent
interactions is different in lighter colliding nuclei compared to
heavier colliding nuclei. In lighter nuclei, the contribution of
mean field towards transverse flow is much smaller compared to
heavier nuclei where binary nucleon-nucleon collisions dominate
the scene. We also find that the inclusion of momentum dependent
interactions explains the energy of vanishing flow in
$^{12}$C+$^{12}$C which, otherwise, was not possible with static
hard equation of state. In this work, we have taken a constant NN
cross-section and stiff equation of state along with its momentum
dependence. The energy and isospin dependence of the cross-section
may also affect the results. It has been shown in \cite {kum98}
that these effects are small in the Fermi energy domain. Further,
the addition of asymmetric potential may alter the equation of
state and balance energy for heavy nuclei, though, it should have
no role for lighter and medium mass colliding nuclei. It should be
further noted that the hard equation of state does not agree with
the experimental data of fragmentation at low energies and
particle production at relativistic energies.

\end{document}